\documentclass[
  aip,
  amsmath,amssymb,
  preprint,
  reprint
]{revtex4-1}

\usepackage{graphicx}
\usepackage{dcolumn}
\usepackage{bm}
\usepackage[utf8]{inputenc}
\usepackage[T1]{fontenc}
\usepackage{mathptmx}
\usepackage{etoolbox}
\usepackage[colorlinks=true, allcolors=blue]{hyperref}
\usepackage{tikz}
\usetikzlibrary{arrows.meta,positioning,shapes.geometric,fit}
\usepackage[caption=false]{subfig} 
\usepackage{multirow}
\usepackage{natbib} 

\begin{document}

\preprint{AIP/123-QED}

\title[Sample title]{Lyapunov Spectral Analysis of Speech Embedding Trajectories in Psychosis}


\author{Jelena Vasi\'c$^{b)}$}
\affiliation{Institute of Mental Health,
Milana Ka\v sanina 3, Belgrade, Serbia}

\author{Branislav Andjeli\'c}
\affiliation{Faculty of Technical Sciences, University of Novi Sad,
Trg Dositeja Obradovica 6, Novi Sad, Serbia}

\author{Ana Man\v ci\'c}
\affiliation{Faculty of Sciences and Mathematics, University of Ni\v s,
Vi\v segradska 33, Ni\v s, Serbia}

\author{Du\v sica Filipovi\'c Djurdjevi\'c}
\affiliation{Faculty of Philosophy, University of Belgrade,
18-20 \v Cika Ljubina street, Belgrade, Serbia}

\author{Ljiljana Mihi\'c}
\affiliation{Faculty of Philosophy, University of Novi Sad,
Dr Zorana Djindji\'ca 2, Novi Sad, Serbia}

\author{Aleksandar Kova\v cevi\'c}
\affiliation{Faculty of Technical Sciences, University of Novi Sad,
Trg Dositeja Obradovica 6, Novi Sad, Serbia}

\author{Nadja P. Mari\'c}
\affiliation{Institute of Mental Health,
Milana Ka\v sanina 3, Belgrade, Serbia}
\affiliation{Faculty of Medicine, University of Belgrade,
Dr Suboti\'ca 8, Belgrade, Serbia}

\author{Aleksandra Maluckov$^{a)}$}
\affiliation{Vin\v{c}a Institute of Nuclear Sciences,
National Institute of the Republic of Serbia,
University of Belgrade, Mike Petrovi\'ca Alasa 12-14, Belgrade, Serbia}


\email[Corresponding author: ]{sandram@vin.bg.ac.rs}     
\email[$^{b)}$ Corresponding author: ]{jelenavas9@gmail.com}   

\date{\today}
\begin{abstract}
We analyze speech embeddings from structured clinical interviews of psychotic patients and healthy controls by treating language production as a high-dimensional dynamical process. 
Lyapunov exponent (LE) spectra are computed from word-level and answer-level embeddings generated by two distinct large language models, allowing us to assess the stability of the conclusions with respect to different embedding presentations. Word-level embeddings exhibit uniformly contracting dynamics with no positive LE, while answer-level embeddings, in spite of the overall contraction, display a number of positive LEs and higher-dimensional attractors. The resulting LE spectra robustly separate psychotic from healthy speech, while differentiation within the psychotic group is not statistically significant overall, despite a tendency of the most severe cases to occupy distinct dynamical regimes. These findings indicate that nonlinear dynamical invariants of speech embeddings provide a physics-inspired probe of disordered cognition whose conclusions remain stable across embedding models.

\end{abstract}

\maketitle

\begin{quotation}
\noindent

We investigate spoken language as a complex dynamical system by analyzing speech embeddings extracted from structured clinical interviews of psychotic patients and healthy controls. Motivated by the universality of complexity across physical, biological, and cognitive systems, we interpret speech production as a trajectory evolving in a high-dimensional semantic space. Word-level and answer-level embeddings are treated as dynamical realizations of linguistic organization, and their stability is quantified using Lyapunov exponent (LE) spectra.

To assess the robustness of the conclusions with respect to linguistic representation, the analysis is performed independently on embeddings generated by two distinct large language models. Across both embedding representations, word-level trajectories exhibit uniformly negative Lyapunov exponents, indicating locally stable and contracting dynamics associated with short-range lexical organization. In contrast, answer-level embeddings in spite of the overall contraction display a number of positive Lyapunov exponents and higher-dimensional attractors. This reflects characteristic discourse-level structure over extended semantic contexts. While the geometric realization of the trajectories depends on the embedding model, the qualitative conclusions regarding psychosis and healthy controls remain unchanged.

The resulting LE spectra provide a clear and statistically robust separation between psychotic and healthy subjects that does not primarily encode clinical severity. Differentiation within the psychotic cohort is not significant overall, although the most severe cases tend to populate distinct regions of the dynamical feature space. Psychosis is associated with a reduction of dynamical richness, compressing discourse trajectories into lower-dimensional and less unstable states.

These findings demonstrate that nonlinear dynamical invariants of speech embeddings capture intrinsic organization of thought beyond symptom-based clinical scales. By combining modern language representations with tools from nonlinear dynamics and complexity theory, this work establishes a physics-inspired framework for probing altered cognitive organization in psychosis.

\end{quotation}

\section{Introduction}

Complexity in nature emerges as a profound manifestation of underlying simplicity. From the interactions of quarks, atoms, and molecules to the emergence of ocean waves, galaxies, life, and consciousness itself, complex organization appears across diverse scales. 

Neural firing patterns in the brain, gravitational collapse in astrophysics, and cosmological structure formation are often viewed as manifestations of a common hierarchy of nonlinear dynamics, governed by instability, self-organization, and universality across scales \cite{Nicolis1977,Bak1996,BeggsPlenz2003,Choptuik1993,CrossHohenberg1993}. 
From gravitational waves to the Big Bang, structured patterns arise spontaneously from fundamental dynamics \cite{Kibble1976,Choptuik1993,CrossHohenberg1993}.
Within this hierarchy resides the organization of the human brain, giving rise to thought and language, and in its altered states, to psychosis—suggesting that cognitive phenomena may be probed using the same physics-inspired tools that have proven effective across natural complex systems.

One of the largest coordinated efforts to understand these phenomena, the \textit{Diverse International Scientific Consortium for Research in Thought, Language and Communication in Psychosis (DISCOURSE)}, focuses on the multiple dimensions of language as a window into psychotic processes \cite{discourse}. Our work aligns with this mission by focusing on recorded structured interviews, analyzing how speech dynamics reflect underlying cognitive organization.  

 Language disturbances, related to formal thought disorder, incoherent speech, and semantic disorganization—are hallmark features of psychotic illnesses such as schizophrenia. The structure of patients’ speech, particularly in narrative tasks or responses to structured clinical questions, reveals restricted lexical diversity, diminished coherence, and repetitive or stereotyped patterns that strongly correlate with cognitive deficits and functional outcomes~\cite{AlonsoSanchez2022, Palaniyappan2021, Tan2023}.

Recent advances in computational psychiatry have led to the application of natural language processing (NLP) methods for the precise characterization of psychotic speech patterns \cite{marder2022nlp,Tan2023}. Computational language analysis methods, including semantic, graph-based, and machine-learning approaches, have been applied in relevant studies to quantify speech abnormalities in psychosis \cite{Elvevag2010,mota2012speech,Mota2017,rezaii2019machine,hitczensko2021}.  These studies report associations between language-based markers and clinical features or illness trajectories, and suggest their potential utility for predicting psychosis-related outcomes \cite{Corcoran2018,AlonsoSanchez2022,corcoran2024}.

While feature-based NLP and graph-theoretic coherence measures have advanced the field, less attention has been given to the underlying semantic trajectories of speech as they evolve over time and to their dynamical organization. Modern embedding methods transform linguistic units into high-dimensional continuous numerical vectors, capturing subtle relationships beyond static lexical or syntactic features. Our approach builds on this capability by representing speech from structured interviews at multiple linguistic levels: answers, sentences, words, and subwords, as trajectories in embedding space. Then, we apply nonlinear dynamical systems Lyapunov spectra analysis.
To assess the robustness of the observed dynamical signatures, we analyze embeddings generated by different transformer-based language models.

To our knowledge, no previous study has combined layered embedding representations with dynamical complexity measures to probe alterations in speech dynamics associated with psychosis and speech in general. By analyzing embedding-based trajectories extracted from structured clinical interviews of psychotic patients and healthy controls, we aim to bridge computational linguistics and complexity science. 
The present study is intentionally exploratory in nature. Rather than aiming at immediate clinical generalization, our goal is to probe whether nonlinear dynamical concepts—specifically Lyapunov spectra and attractor dimensionality—can reveal meaningful structure in discourse embeddings of psychiatric interviews. We seek to stimulate further investigation into dynamical approaches to language-based psychopathology across larger cohorts, diverse languages, and alternative embedding representations.

In this manuscript, Section II introduces our hierarchical embedding methodology, implemented across multiple linguistic layers using transformer-based models adapted for Serbian (BERTi\'c \cite{ljubesic2021bertic} and Multilingual E5 \cite{wang2024multilingual}). Section III presents a dynamical systems analysis of embedding trajectories, computing Lyapunov exponents and related measures to assess complexity differences between psychotic and control groups. The results are reported and discussed in Section IV, while Section V concludes the study. Appendix contains results which confirm the main findings.

\section{Multilevel embedding and dynamical analysis of structured psychiatric interviews}
\label{sec:multilayer_embeddings}

The temporal structure of human speech reflects a high-dimensional, nonlinear dynamical process shaped by cognition, affect, motivation, and mental state. In structured interviews used for psychosis assessment, this process unfolds across multiple linguistic levels: \textit{answers} (entire responses to a psychiatrist's question), \textit{sentences} (subdivisions of answers), \textit{words}, and potentially \textit{sublexical}. Each level captures a different granularity of information: from coarse semantic organization at the answer level to fine-grained lexical or sublexical patterns. 

Our approach aims to extract dynamical information from high-dimensional embedding vectors derived from these levels. Unlike static linguistic analysis \cite{bedi2015predicting}, we conceptualize embeddings as constituents of trajectories in a semantic state-space, where the evolution over time reflects the underlying cognitive dynamics of the subject. The central hypothesis is that psychosis manifests not only through changes in lexical choice or sentence structure but through altered dynamical organization of speech potentially detectable via nonlinear measures such as Lyapunov exponents, fractal dimensions, and multifractal spectra \cite{todder2013chaotic,ma2021fractal,argolo2024rqa}.

\subsection{Clinical Study Protocol for Psychosis}
 
To systematically collect speech data in psychosis, we adopt the standardized \textit{DISCOURSE in Psychosis Speech Bank Protocol}. 
This protocol \cite{protocol} is specifically designed for clinical and research settings to elicit naturalistic yet structured speech samples from individuals diagnosed with psychotic disorders. 

Recordings are performed with calibrated microphones in routine healthcare environments (Fig. \ref{sheme}). The protocol lasts approximately $20$ minutes and requires minimal interviewer intervention to ensure spontaneous responses. It comprises seven task modules: free conversation (brief open-ended dialogue on daily life and interests); personal and health narratives (first-person stories about life events and mental health history); picture description (participants describe and interpret standardized images); storyboard retelling (reconstructing a visual story sequence to assess narrative coherence); dream reporting (recalling recent or recurrent dreams to elicit imaginative discourse), reading aloud and recall (reading a short story followed by free recall to assess memory and articulation).

In addition, demographic, clinical, and medication data are collected, including PANSS ratings of symptom severity. 
This structured yet flexible design captures multiple speech modalities (spontaneous, descriptive, narrative, and recall), 
allowing for comprehensive linguistic and psycholinguistic analysis across different stages and severities of psychosis.

\subsection*{Ethical approval}
The study protocol was approved by the Ethical Committee of the Institute of Mental Health (approval number 1060/2058/1) and the Institutional Review Board of the Department of Psychology, Faculty of Philosophy, University of Belgrade, Serbia ($\# 2023-78$). All procedures were conducted in accordance with the Declaration of Helsinki. Written informed consent was obtained from all participants prior to participation in the study.

\subsection{Clinical characteristics of the psychotic cohort and implications for dynamical analysis}

The psychotic cohort analyzed in this study consists of $27$ individuals with psychosis (18 males and 9 females; mean age $28.9\pm 8.1$ years) which reflects the heterogeneity typical of real-world psychiatric populations. Diagnoses span schizophrenia spectrum disorders (ICD-10 codes F20.x), acute and transient psychotic disorders (F23.x), schizoaffective and related disorders (F25.x), and other non-organic psychoses (F29). Diagnostic labels were determined by treating psychiatrists as part of routine clinical assessment.

Symptom severity within the cohort shows substantial variability. See Appendix~\ref{app:clinical_cohort} for cohort details. Positive and Negative Syndrome Scale (PANSS) total scores range from approximately 49 to above 110, with diverse contributions from the positive (P), negative (N), and general psychopathology (G) subscales. Importantly, subjects with comparable PANSS total scores may differ markedly in symptom composition, for example exhibiting predominance of positive symptoms (e.g., hallucinations and delusions) versus negative symptoms (e.g., affective flattening, alogia, or cognitive slowing). This heterogeneity indicates that PANSS total should be understood as a coarse measure of overall symptom burden rather than a unidimensional severity axis.

In the present draft, PANSS total score is used pragmatically to stratify subjects into severity groups for group-level statistical analysis: Mild, Moderate, and Severe groups (Fig. \ref{sheme}), while full clinical characteristics of the cohort are reported in Appendix~\ref{app:clinical_cohort}.
This choice reflects clinical practice and allows comparison with prior computational psychiatry studies, while acknowledging that it does not capture the full multidimensional structure of psychopathology. Consequently, group labels should be interpreted as approximate organizational categories rather than strict dynamical classes.

For reference, the control group consists of 10 healthy subjects with no history of psychotic or major psychiatric disorders  (6 males and 4 females; mean age $30.5 \pm 10.3$ years). These participants underwent the same structured interview protocol and embedding pipeline as the patient cohort, providing a normative baseline for discourse-level dynamical analysis. This design enables direct comparison of speech dynamics between healthy and psychotic populations under identical methodological conditions.

\subsection{Speech parameters as reference indicators of psychosis}

Language disturbances are a core and persistent feature of psychotic disorders, particularly schizophrenia, and have long been recognized as objective indicators of altered cognitive organization. Clinical psychiatry has traditionally described these disturbances in terms of formal thought disorder, semantic disorganization, and incoherent speech has long been described. Over the past two decades, these qualitative descriptions have been progressively formalized using computational approaches, establishing a set of speech-derived parameters that now serve as reference indicators in the study of psychosis
\cite{Elvevag2007,mota2012speech,Mota2017,rezaii2019machine,hitczensko2021,AlonsoSanchez2022,Palaniyappan2021}.

One of the most consistently reported findings concerns semantic coherence and similarity. Psychotic speech often exhibits increased local word-to-word similarity accompanied by reduced global coherence across longer discourse segments. Embedding-based approaches have shown that persistent semantic proximity between successive words or phrases reflects constrained semantic exploration and is associated with negative symptoms and chronic illness trajectories
\cite{Elvevag2007,rezaii2019machine,hitczensko2021}. Closely related to this is the concept of semantic density, where psychotic discourse may remain fluent yet convey reduced informational content. Lower semantic density has been shown to predict psychosis onset and transition with high accuracy in longitudinal and high-risk cohorts
\cite{rezaii2019machine,AlonsoSanchez2022}.

Beyond local semantic measures, graph-theoretic representations of speech have revealed structural disorganization at the level of discourse architecture. By representing speech as networks of connected lexical units, studies have demonstrated that graph metrics such as reduced connectivity, altered path lengths, and increased recurrence capture disorganized thought patterns and reliably differentiate schizophrenia from mood-related psychoses
\cite{mota2012speech,Mota2017}. These findings emphasize that psychoses-related language abnormalities are not merely lexical but reflect altered global organization.

Lexical and syntactic anomalies provide further evidence of disrupted linguistic structure. Reduced lexical diversity, atypical use of pronouns and function words, and simplified syntactic constructions have been repeatedly reported using tools such
as the Linguistic Inquiry and Word Count (LIWC) framework and syntactic parsing approaches
\cite{Palaniyappan2021}. These features reflect constraints on expressive flexibility and grammatical organization rather than isolated vocabulary deficits.

Finally, alterations in prosody and fluency constitute an additional, partially independent dimension of psychotic speech. Flattened affect, prolonged pauses, reduced pitch variability, and altered speech rate are frequently observed and can be quantified through acoustic analysis
\cite{hitczensko2021}. While these features are often linked to symptom severity, they also reflect broader disruptions in temporal coordination and expressive dynamics.

\begin{figure}[t]
\centering
\begin{tikzpicture}[
    node distance=1.7cm,
    every node/.style={draw, rectangle, rounded corners, align=center, minimum width=3.6cm, minimum height=0.9cm},
    arrow/.style={->, thick}
]

\node (interview) {Structured clinical interviews\\(audio recordings)};
\node (transcript) [below of=interview] {Transcription by ASR\\  + manual correction (patient answers)};
\node (embedding) [below of=transcript] {LLM embeddings\\(word /sentence / answer)};
\node (panss) [below of=embedding] {Clinical grouping by PANSS total:\\ Mild: 58 $<$ PANSS $\leq$ 74\\
 Moderate: 74 $<$ PANSS $\leq$ 94\\
    Severe : PANSS $\geq$ 95};
\node (lyap) [below of=panss] {Lyapunov spectrum\\computation: Dynamical features\\$LE_{\max}, \sum LE, N_{pos}, D_{KY}$ };
\node (stats) [below of=lyap] {Statistical analysis\\(ANOVA, KW, effect sizes)};
\node (result) [below of=stats] {Dynamical separation\\of clinical groups};
\draw[arrow] (interview) -- (transcript);
\draw[arrow] (transcript) -- (embedding);
\draw[arrow] (embedding) -- (panss);
\draw[arrow] (panss) -- (lyap);
\draw[arrow] (lyap) -- (stats);
\draw[arrow] (stats) -- (result);

\end{tikzpicture}
\caption{Schematic overview of the analysis procedure.}
\label{sheme}
\end{figure}
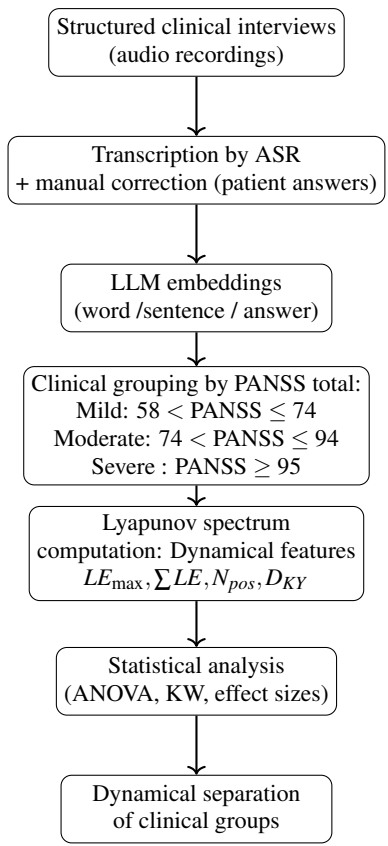

Taken together, these speech parameters define a well-established reference framework for characterizing psychotic language. Importantly, most existing approaches quantify these features as static descriptors or summary statistics. In contrast, the present work builds on these findings by treating speech embeddings as constituents of evolving trajectories and probing their organization using tools from nonlinear dynamical systems theory, thereby shifting the focus from isolated features to the dynamical structure of linguistic processes (Fig. \ref{sheme}).

\subsection{Hierarchical embedding representation}
\label{sec:hierarchical_embedding}

Speech recordings were transcribed using an automatic speech recognition (ASR) system, followed by manual correction (Fig. \ref{sheme}). The transcribed speeches from structured clinical interviews are then represented using hierarchical embedding levels that capture linguistic organization at multiple scales (Fig. \ref{sheme}). Each level is defined by the sequence of vectors forming a trajectory in a high-dimensional semantic space, enabling analysis using tools from nonlinear dynamical systems theory.

We consider the following levels of representation:

\begin{enumerate}
    \item \textit{Answer-level embeddings.} Each subject’s response to a clinical question is encoded as a single vector capturing the global semantic content of the answer. Transformer-based models naturally generate such coarse-grained representations through pooling mechanisms (e.g., CLS token or mean pooling), which are commonly used to summarize long-range semantic context \cite{devlin2019bert,reimers2019sentence}.

   \item \textit{Sentence-level embeddings:} Answers are segmented into sentences, each sentence embedded separately. This layer captures intra-answer organization and potential loss of discourse coherence, which has been quantified in NLP using entity-based and discourse/cohesion models \cite{barzilay2008entitygrid,graesser2004cohmetrix,lin2011discoursecoherence}.

    \item \textit{Word-level embeddings.} Words are embedded sequentially, preserving the temporal order of speech production. This level provides the highest temporal resolution in this study and forms the primary basis for the dynamical trajectory analysis (Appendix \ref{app:word_strategy}).

\end{enumerate}

Embeddings were generated using two independent large language models (Fig. \ref{sheme}). The first is BERTi\'c, a transformer model adapted for the Serbian language, producing embeddings of dimensionality $d=768$ \cite{palominos2024semantic}. The second model (Multilingual E5 \cite{wang2024multilingual}) produces embeddings of dimensionality $d=1024$. Although these models differ in architecture and embedding dimensionality, both generate continuous vector representations that can be interpreted as realizations of high-dimensional dynamical systems.

Importantly, the Lyapunov exponent analysis employed in this work probes the local stability and divergence of trajectories and does not require the embedding spaces to have equal dimensionality. Consistent qualitative conclusions across both embedding models therefore indicate that the observed dynamical signatures are robust to representation choice rather than tied to a specific embedding geometry.

\section{Dynamical analysis of embedding vectors via Lyapunov spectra}
\label{sec:lyapunov}

\subsection{Dynamical interpretation of speech embeddings}
\label{sec:dynamical_interpretation}

Once embedded, each interview (restricted to subject responses) is represented by an embedding matrix
\[
    \mathbf{X} =
    \begin{pmatrix}
    \mathbf{x}(t_0) \\
    \mathbf{x}(t_1) \\
    \vdots \\
    \mathbf{x}(t_{T-1})
    \end{pmatrix}
    \in \mathbb{R}^{T \times N},
\]
where each row $\mathbf{x}(t) = (x_1(t), x_2(t), \dots, x_N(t))^\top$ is a generalized semantic vector in a high-dimensional space, with $N = 768$ or $1024$ depending on the embedding model. Each component $x_i(t)$ carries some information related to the state of the system within the N-dimensional state at the given moment in time, which we hypothesise is in turn related to the subject’s cognitive, emotional, and linguistic state at a given moment of speech production.

Beyond this layer lies the deep structure of the transformer-based language model—a nonlinear learned mapping from linguistic tokens to semantic coordinates. This mapping is shaped stochastically during training and remains largely opaque to direct interpretation. 

However, the physiological process in psychosis speeches that generates the embeddings is inherently dynamic. To bridge this gap, we reinterpret the embedding matrix as a discrete-time realization of a hidden dynamical system governing thought and language production during the interview. In this perspective, each embedding dimension $x_i(t)$ represents the temporal evolution of a generalized coordinate of an underlying cognitive--linguistic system, and the full embedding vector evolves according to
\[
    \mathbf{x}(t+1) = \mathbf{F}(\mathbf{x}(t)) + \boldsymbol{\xi}(t),
\]
where $\mathbf{F}: \mathbb{R}^N \rightarrow \mathbb{R}^N$ is an unknown nonlinear map implicitly shaped by the subject’s cognitive processes and the geometry of the embedding space, while $\boldsymbol{\xi}(t)$ represents stochastic fluctuations.

At the initial time $t=0$, the interview is characterized by a configuration of $N$ coupled trajectories encoding the semantic--cognitive state of the subject. As the interview progresses ($t = 1, 2, \dots, T-1$), these trajectories evolve on a high-dimensional manifold jointly determined by the subject’s mental organization and the embedding model. Each linguistic unit—whether corresponding to a full answer, a sentence, or a word—thus constitutes a stroboscopic snapshot of this latent dynamical system.

This reinterpretation naturally shifts the focus from static semantic similarity to questions of stability, sensitivity to perturbations, and effective dimensionality of speech dynamics.

\subsection{Lyapunov spectrum and stability analysis}
\label{sec:lyapunov_theory}

To characterize the stability and complexity of the embedding trajectories, we compute the full Lyapunov spectrum, following foundational approaches in nonlinear dynamical systems theory
\cite{strogatz2018,lichtenberg1992,eckmann1985} (Fig. \ref{sheme}). Considering infinitesimal perturbations $\delta \mathbf{x}(t)$ along orthogonal directions in embedding space, the tangent dynamics are governed by
\[
    \delta \mathbf{x}(t+1) = \mathbf{J}(t)\,\delta \mathbf{x}(t),
\]
where $\mathbf{J}(t) = \frac{\partial \mathbf{F}}{\partial \mathbf{x}} \big|_{\mathbf{x}(t)}$ denotes the Jacobian of the effective dynamical map evaluated along the trajectory.

The Lyapunov exponents are defined as
\[
    \lambda_i = \lim_{n \to \infty} \frac{1}{n} 
    \sum_{k=0}^{n-1}
    \ln \frac{\|\delta \mathbf{x}_i(k+1)\|}{\|\delta \mathbf{x}_i(k)\|},
    \qquad i = 1, \dots, N,
\]
and quantify the average exponential rates of divergence or contraction of nearby trajectories. Positive Lyapunov exponents indicate sensitivity to perturbations and increased dynamical complexity related to divergence of nearby trajectories, whereas negative exponents correspond to contracting dynamics.

In practice, the full Lyapunov spectrum is estimated using a QR-based implementation of the Benettin algorithm with periodic orthonormalization to ensure numerical stability
\cite{benettin1980,benettin1980b,wolf1985}. This approach tracks the evolution of an orthogonal set of perturbation vectors along the embedding trajectory, yielding the complete spectrum of exponents.

From the computed Lyapunov spectrum, we extract several summary quantities relevant for characterizing linguistic dynamics:
\begin{itemize}
    \item the maximum Lyapunov exponent ($LE_{max}$), capturing the dominant local expansion or contraction rate;
    \item the sum of Lyapunov exponents ($\Sigma$), reflecting phase-space volume contraction;
    \item the number of positive Lyapunov exponents ($N_{LE}$), indicating the effective degree of instability;
    \item the Kaplan--Yorke (KY) fractal dimension,
    \[
        D_{KY} = j + \frac{\sum_{i=1}^{j} \lambda_i}{|\lambda_{j+1}|},
        \quad \text{where } \sum_{i=1}^{j} \lambda_i \ge 0,
    \]
    which estimates the dimensionality of the underlying attractor.
\end{itemize}

Within this framework, psychosis may manifest as altered stability and reduced effective dimensionality of the semantic attractor underlying speech. We therefore hypothesize that psychotic speech dynamics is characterized by (i) reduced maximum Lyapunov exponents and fewer positive exponents, reflecting diminished sensitivity to semantic perturbations; (ii) lower Kaplan--Yorke dimension, indicating contraction of the accessible semantic state space; and (iii) changes in phase-space volume contraction, potentially associated with rigid or stereotyped linguistic patterns.

\section{Results and Discussion}
\label{sec:results}

\subsection{Micro--macro transition in speech dynamics: word versus answer embeddings}
\label{sec:micro_macro}

We first examined the dynamical properties of speech embeddings at the word level, treating sentence-context word embeddings as components of high-dimensional trajectories (Appendix \ref{app:word_strategy}). Across all clinical groups, the Lyapunov spectra were strictly negative, with no positive Lyapunov exponents observed. This indicates uniformly contracting, dissipative dynamics, consistent with stable local semantic evolution. The absence of positive exponents implies that word-level embeddings do not exhibit sensitive dependence on initial conditions, regardless of clinical status.

\begin{table}[h]
\centering
\caption{Lyapunov spectrum statistics for word-level sentence embeddings.}
\label{tab:word_level}
\begin{tabular}{|l|c|c|c|}
\hline
Group & $\langle LE_{\max}\rangle \pm \sigma$ & $<\Sigma> \pm \sigma(\Sigma)$ & $N_{LE}$ \\
\hline
Severe   & $-0.190 \pm 0.031$ & $-269.13 \pm 13.80$ & $0$ \\
Moderate & $-0.226 \pm 0.047$ & $-295.29 \pm 41.31$ & $0$ \\
Mild     & $-0.256 \pm 0.052$ & $-319.58 \pm 41.37$ & $0$ \\
Healthy  & $-0.199 \pm 0.028$ & $-264.53 \pm 17.35$ & $0$ \\
\hline
\end{tabular}
\end{table}

Although quantitative differences in the magnitude of the Lyapunov exponents are present across groups, these variations do not induce qualitative changes in the dynamical regime. In particular, the Moderate and Mild psychosis groups exhibit the most negative maximum and total Lyapunov exponents, indicative of reduced local semantic variability and flattened lexical dynamics. Overall,  word-level embeddings behave as a localized, mean-field representation of speech, capturing moment-to-moment semantic coherence while lacking dynamical richness. 

Sentence-level embeddings exhibit strongly heterogeneous Lyapunov spectra within each
clinical group, with mixtures of fully contracting and weakly expanding dynamics,
resulting in large within-group variance (Appendix~\ref{app:sentence_lyap}).

\begin{table*}[t]
\centering
\caption{Group-averaged Lyapunov spectrum characteristics for BERTi\'c  answer embeddings under four-group clinical classification $<...>$.
All quantities are reported as mean $\pm$ standard deviation across subjects within each group.}
\label{tab:bertic_answer_4groups}
\begin{tabular}{|l|c|c|c|c|}
\hline
Group & $<LE_{max}> $ & $<\Sigma>$ &  $<N_{LE}>$ &
$<D_{KY}>$ \\
\hline
Severe   & $0.189 \pm 0.053$ & $-260.27 \pm 29.41$ & $25.86 \pm 15.60$ & $66.92 \pm 38.46$ \\
Moderate & $0.238 \pm 0.070$ & $-284.99 \pm 49.14$ & $34.30 \pm 26.31$ & $83.54 \pm 63.75$ \\
Mild     & $0.144 \pm 0.084$ & $-314.00 \pm 63.26$ & $26.11 \pm 28.55$ & $63.14 \pm 68.41$ \\
Healthy  & $0.341 \pm 0.080$ & $-292.94 \pm 42.00$ & $63.00 \pm 31.97$ & $146.15 \pm 71.42$ \\
\hline
\end{tabular}
\end{table*}

A qualitatively different regime emerges when analyzing answer-level embeddings.
In contrast to the strictly contracting word-level dynamics, Lyapunov spectrum statistics for answer embeddings derived from the BERTi\'c model (Table~\ref{tab:bertic_answer_4groups}) reveal a qualitatively different dynamical regime.

In contrast to the word-level results, answer embeddings exhibit positive maximal Lyapunov exponents and multiple positive dimensions across all clinical and control groups, indicating instability and increased dynamical complexity at the discourse scale. This transition from fully contracting dynamics at the word level to partially expanding dynamics at the answer level reflects a clear micro--macro transition in semantic organization.

This divergence can be interpreted through scaling and semantic aggregation. Word embeddings operate at a micro-level, capturing localized semantic units within individual sentences. Because sentences typically remain semantically tight—even in psychotic speech—the resulting trajectories are smooth and dissipative. In contrast, answer embeddings integrate entire patient responses, often spanning multiple sentences, thematic shifts, and emotional modulation, into a coarse-grained semantic representation. The increased variability of these units enables the emergence of positive Lyapunov directions: trajectories in embedding space become sensitive to perturbations and explore higher-dimensional regions. In healthy speech, this manifests as richer semantic divergence, whereas more severe psychosis diminishes this diversity, reducing both the number and magnitude of positive exponents.

These findings align with established observations in psychiatry and computational linguistics. Disruptions in psychosis manifest predominantly at the discourse level—through derailment, tangentiality, and thematic drift—rather than within individual sentences \cite{mota2012speech,Mota2017,bedi2015predicting,Corcoran2018}. Graph-based representations of speech capture this macroscopic fragmentation via reduced connectedness and increased recurrence, whereas local sentence coherence often remains preserved \cite{Mota2018,Palaniyappan2020}. Similarly, studies of temporal semantic trajectories emphasize that shifts across utterances are more revealing of psychosis than static lexical averages or short-range coherence measures \cite{Elvevag2010}.

From a dynamical system perspective, healthy discourse corresponds to exploration of a higher-dimensional semantic attractor with multiple unstable directions, whereas psychotic speech tends to collapse into lower-dimensional, more stable attractors. This contraction is consistent with reduced semantic flexibility and cognitive narrowing reported in psychosis \cite{Durstewitz2018,Stam2021}. These results indicate that word-level dynamics remain stable and fully dissipative across all groups. On the other hand,  answer embeddings and their Lyapunov spectra capture clinically meaningful complexity and organizational structure
that remain invisible at the sentence micro-level. Taken together, this motivates the analysis of larger semantic units where discourse-level structure,
flexibility, and cognitive variability may emerge.

\subsection{Answer-level dynamics distinguish psychotic and healthy speech}
\label{sec:answer_results}

Building on the findings reported above, we  want to investigate whether discourse-level dynamical properties can differentiate between healthy controls and psychosis related group. Also, we explore how these properties differ across different clinical groups. All this was done using full Lyapunov spectra and Kaplan-York (KY) dimensions computed from answer-level embeddings. 

Across all clinical groups, the maximal Lyapunov exponent is positive, indicating sensitivity to perturbations and nontrivial nonlinear dynamics at the discourse scale. 
However, substantial quantitative differences emerge between healthy and psychotic subjects in higher-order dynamical characteristics.

Healthy participants exhibit a markedly larger number of positive Lyapunov exponents and substantially higher Kaplan--Yorke dimensions, reflecting richer and higher-dimensional attractor structures. In contrast, psychotic speech is characterized by fewer positive exponents and lower KY dimensions, consistent with contraction of the accessible semantic state space and reduced dynamical degrees of freedom. These differences point to a reduced capacity for semantic exploration and thematic diversification in psychotic discourse. Table \ref{tab:bertic_answer_2groups} summarizes the LE statistics for Healthy vs. Psychotic groups, while Table \ref{tab:bertic_answer_3groups} makes more subtle differentiation to Healthy, Severe and Mild+Moderate groups.

\begin{table*}[t]
\centering
\caption{LE spectrum statistics for BERTi\'c answer embeddings under two-group classification (Healthy vs. Psychosis). Values are given as mean $\pm$ standard deviation.}
\label{tab:bertic_answer_2groups}
\begin{tabular}{|l|c|c|c|c|}
\hline
Group & $<LE_{max}> $ & $<\Sigma>$ &  $<N_{LE}>$ & $<D_{KY}>$ \\
\hline
Psychosis & $0.192 \pm 0.080$ & $-288.38 \pm 54.69$ & $29.19 \pm 24.20$ & $72.00 \pm 58.35$ \\
Healthy   & $0.341 \pm 0.080$ & $-292.94 \pm 42.00$ & $63.00 \pm 31.97$ & $146.15 \pm 71.42$ \\
\hline
\end{tabular}
\end{table*}

\begin{table*}[t]
\centering
\caption{LE spectrum statistics for BERTi\'c answer embeddings under three-group classification. Values are given as mean $\pm$ standard deviation.}
\label{tab:bertic_answer_3groups}
\begin{tabular}{|l|c|c|c|c|}
\hline
Group & $<LE_{max}> $ & $<\Sigma>$ &  $<N_{LE}>$ & $<D_{KY}>$ \\
\hline
Severe   & $0.189 \pm 0.053$ & $-260.27 \pm 29.41$ & $25.86 \pm 15.60$ & $66.92 \pm 38.46$ \\
Mild+Moderate    & $0.193 \pm 0.089$ & $-298.73 \pm 58.11$ & $30.42 \pm 26.95$ & $73.87 \pm 64.97$ \\
Healthy  & $0.341 \pm 0.080$ & $-292.94 \pm 42.00$ & $63.00 \pm 31.97$ & $146.15 \pm 71.42$ \\
\hline
\end{tabular}
\end{table*}

\begin{figure*}[t]
    \centering
    \includegraphics[width=0.92\linewidth]{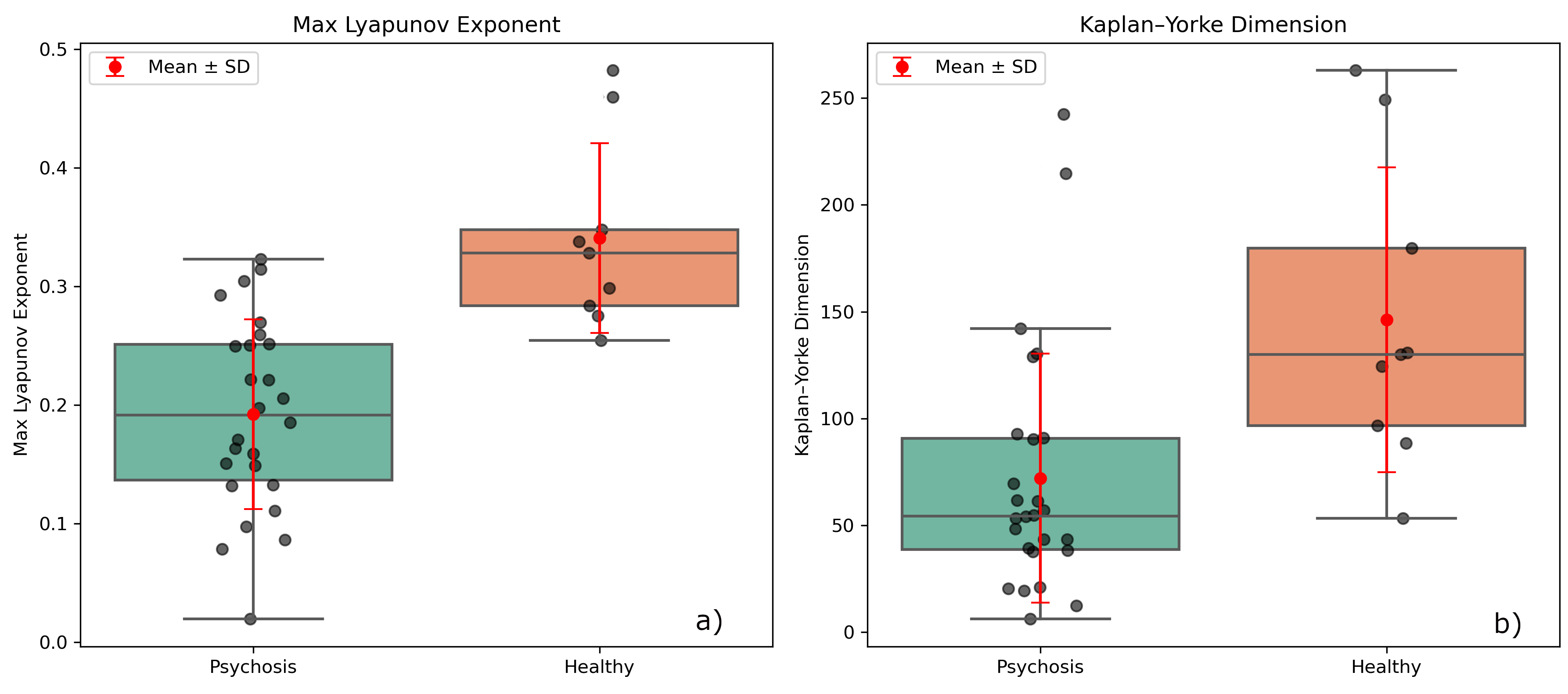}
    \caption{Box plots and corresponding mean values of the maximal Lyapunov exponent and
     the Kaplan--Yorke dimension for answer embeddings in healthy controls
    and subjects with psychosis.}
    \label{fig:answer_2group_lyap}
\end{figure*}

Beyond descriptive differences in attractor dimensionality, the separation between healthy and psychotic discourse is already evident at the level of the maximal Lyapunov exponent, Fig.~\ref{fig:answer_2group_lyap}(a), and the Kaplan--Yorke dimension, Fig.~\ref{fig:answer_2group_lyap}(b). A direct two-group statistical comparison (Psychosis vs.\ Healthy) reveals a highly significant reduction of the maximal Lyapunov exponent in psychotic speech. Using Welch’s unequal-variance $t$-test, the group difference is strongly significant ($t = -4.81$, $p = 2.8 \times 10^{-4}$), and this result is independently confirmed by a nonparametric Mann--Whitney $U$ test ($p = 1.7 \times 10^{-4}$) as well as a distribution-free permutation test ($p \approx 5 \times 10^{-5}$).

Effect sizes were quantified using both Cohen’s $d$ and Cliff’s $\delta$, which are robust to unequal variances and non-Gaussian distributions and are recommended for small to moderate sample sizes \cite{Cliff1993,Delaney2002,Lakens2013}. The associated effect sizes are large to extreme (Cohen’s $d = -1.86$, Cliff’s $\delta = -0.85$), indicating that the observed reduction in maximal Lyapunov exponent represents a robust and structurally meaningful difference rather than a marginal statistical effect. Normality tests do not indicate strong deviations from Gaussianity, although all inferences are corroborated by nonparametric and permutation-based analyses.

From a dynamical systems perspective, this result implies that psychotic discourse exhibits reduced sensitivity to semantic perturbations even along the dominant expanding direction. While both healthy and psychotic speech retain positive maximal Lyapunov exponents—signaling nontrivial, nonlinear dynamics—the systematically lower values observed in psychosis suggest diminished semantic flexibility and a tendency toward more stable, constrained discourse trajectories. Importantly, this separation emerges at the discourse scale and is not observed at the sentence or word level, reinforcing the interpretation that psychosis-related alterations in speech dynamics primarily manifest through macroscopic organization rather than local linguistic coherence.

\begin{figure*}[t]
    \centering
    \includegraphics[width=0.92\linewidth]{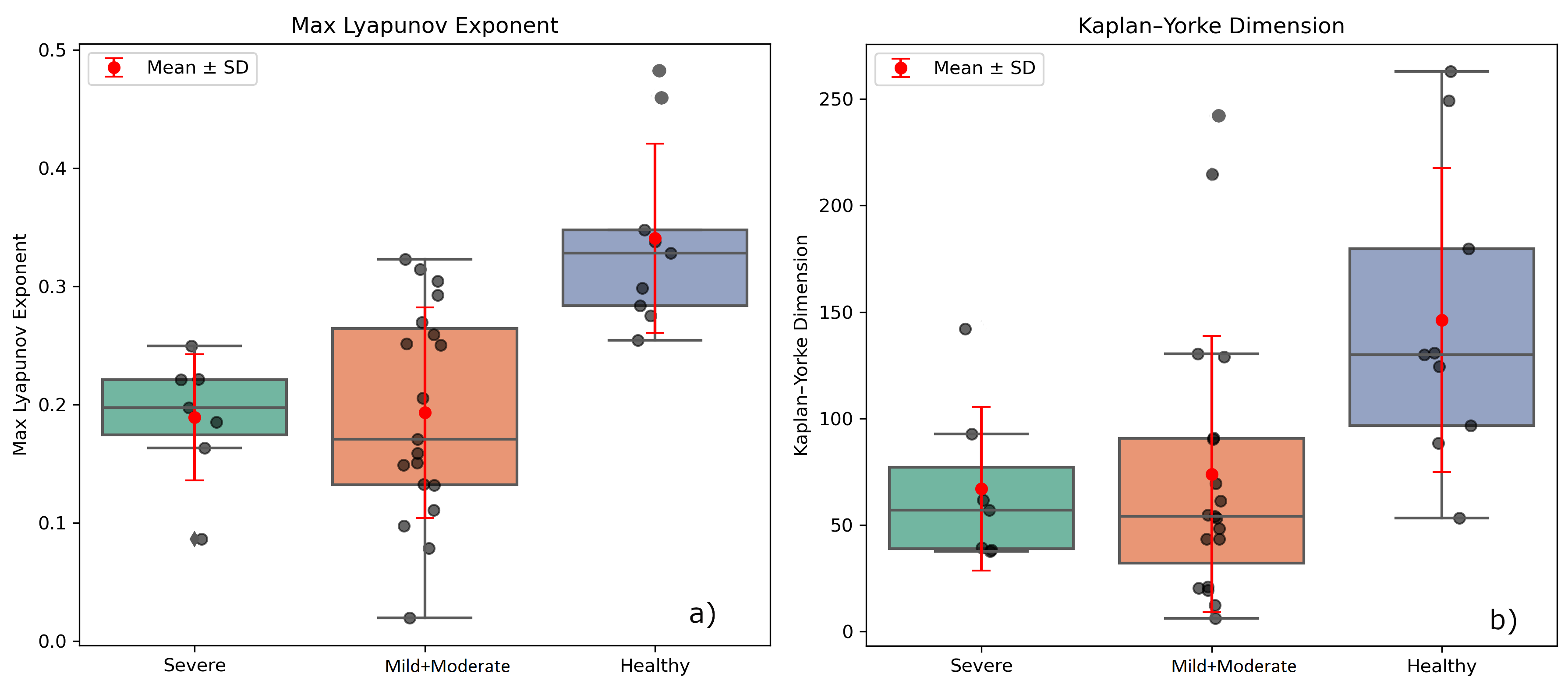}
    \caption{Box plots and corresponding mean values of the maximal Lyapunov exponent and
     the Kaplan--Yorke dimension for answer embeddings for subjects stratified in 3 groups: healthy, mild+moderate and severe psychosis.}
    \label{fig:answer_3group_lyap}
\end{figure*}

We next examined whether the observed separation exhibits a continuous dependence on clinical severity or instead indicates a categorical structural difference in discourse dynamics. When psychotic subjects are stratified into three groups (Severe, Mild+Moderate, Healthy) Fig. \ref{fig:answer_3group_lyap}, a significant overall group effect is observed for the maximal Lyapunov exponent (one-way ANOVA: $F = 11.20$, $p = 2.1 \times 10^{-4}$; Kruskal--Wallis: $H = 14.34$, $p = 7.7 \times 10^{-4}$). However, pairwise comparisons reveal that this effect is driven almost entirely by differences between healthy subjects and both psychotic subgroups.

Specifically, no significant difference is observed between Severe and Mild+Moderate psychosis (Holm-corrected $p > 0.8$, negligible effect size), whereas both groups differ strongly from healthy controls (Holm-corrected $p \approx 10^{-3}$, with large to extreme effect sizes). This absence of separation within the psychotic cohort indicates that Lyapunov-based measures do not map monotonically onto symptom severity, but instead capture a structural distinction between healthy and psychotic discourse dynamics.

Consistent with the micro--macro transition identified above, these findings clarify that psychosis-related alterations in speech dynamics emerge primarily at the discourse scale rather than at the level of local sentence structure. While sentence-level dynamics remain stable across groups, answer-level embeddings reveal marked differences in dynamical organization.

Overall, healthy discourse explores a higher-dimensional semantic attractor with multiple unstable directions, whereas psychotic speech evolves within lower-dimensional, more constrained attractors. In severe psychosis this contraction is most pronounced, reflecting reduced semantic flexibility and diminished exploration of the embedding space. Lyapunov spectra thus capture structural properties of discourse dynamics—such as flexibility, instability, and dimensionality—that are complementary to conventional symptom-based assessments and inaccessible to static or purely local linguistic measures.

\subsubsection{Inter-subject variability and outlier behavior.}
While group-level trends are robust, analysis of individual maximal Lyapunov exponents reveals substantial inter-subject variability within psychotic severity classes. In particular, some subjects clinically classified as Severe or Moderate exhibit maximal Lyapunov exponents comparable to those observed in milder cases, whereas a subset of Mild subjects displays values closer to the lower tail of the Healthy distribution. Such overlaps are expected in heterogeneous psychiatric populations and reflect the fact that severity categories, based on total PANSS scores, aggregate multiple symptom dimensions rather than defining a single dynamical axis.

Importantly, these deviations do not undermine the primary result: the separation between Healthy and Psychosis remains statistically strong and robust across all analyses. Instead, the presence of outliers highlights that Lyapunov-based measures capture intrinsic dynamical organization of discourse, which may vary independently of coarse clinical labels. From a dynamical systems perspective, psychotic speech appears to occupy a family of overlapping attractors with variable dimensionality, rather than a single severity-ordered manifold. This observation supports the interpretation of Lyapunov spectra as structural descriptors of speech dynamics, complementary to—but not reducible to—symptom severity scales.

\subsection{Robustness across language models}
\label{sec:robustness}

To assess the robustness of nonlinear dynamical markers with respect to the choice of language model, we repeated the Lyapunov analysis using an alternative transformer-based embedding model (Multilingual E5 \cite{wang2024multilingual}; $d=1024$). A compact comparison of key Lyapunov-related quantities for answer embeddings and full statistical results are reported in the Appendix.

Both language models yield qualitatively consistent patterns. In both cases, healthy speech exhibits higher values of $LE_{max}$, higher KY dimensions and a larger number of positive Lyapunov exponents, whereas psychotic speech evolves in lower-dimensional attractors. Differences between models are modest and primarily affect the absolute scale of the estimated quantities rather than the relative ordering of clinical groups. These results indicate that the observed dynamical signatures reflect intrinsic structural properties of speech dynamics rather than artifacts of a particular embedding model.


 \section{Summary}

In this work, we used Lyapunov spectra to characterize speech embeddings derived from structured clinical interviews as components of trajectories of an underlying dynamical system. Within this framework, positive Lyapunov exponents indicate sensitivity to semantic perturbations and flexible exploration of meaning, while negative spectra correspond to contracting, rigid dynamics. The KY dimension, derived from the full spectrum, provides an estimate of the effective number of dynamical degrees of freedom available to speech evolution.

Our results reveal a pronounced scale dependence in the dynamical organization of language. At the micro scale, sentence-level word embeddings exhibit strictly negative Lyapunov spectra across all clinical groups. This uniform contraction indicates that local semantic dynamics remain stable and tightly constrained, even in psychotic speech. Clinically, this aligns with the observation that formal thought disorder rarely manifests as a breakdown of sentence-level coherence, but instead emerges at broader discourse scales.

At the macro scale, answer-level embeddings display qualitatively different behavior. Full patient responses exhibit positive maximal Lyapunov exponents and multiple unstable directions, indicating richer and more complex dynamics. Healthy subjects consistently occupy higher-dimensional semantic attractors, while psychotic speech evolves within lower-dimensional, more constrained regions of embedding space. Importantly, this reduction in dynamical dimensionality reflects a loss of semantic flexibility rather than a simple increase in noise or instability.

Although the overall dynamics remain globally dissipative for all groups—as reflected by strongly negative sums of Lyapunov exponents—the internal structure of the spectrum differs markedly between healthy and psychotic speech. It is the distribution of unstable directions and the resulting attractor dimensionality, rather than the maximal Lyapunov exponent alone, that provides the clearest separation between groups. Notably, these dynamical measures do not map monotonically onto clinical severity scales, indicating that Lyapunov spectra probe structural organization of discourse rather than symptom intensity per se.

In contrast, static embedding-based distance measures showed limited discriminatory power between clinical groups (\ref{app:vidi}). This highlights the necessity of temporal and dynamical analysis: clinically meaningful information about thought disorder resides not in isolated semantic states, but in how speech trajectories evolve over time.

Taken together, our findings support a picture in which healthy discourse explores a high-dimensional semantic state space with multiple unstable directions, resembling a dynamically rich system capable of flexible thematic transitions. Psychotic speech, by contrast, collapses toward lower-dimensional attractors, consistent with cognitive narrowing and reduced adaptability. In this sense, Lyapunov-based invariants provide physics-inspired markers of discourse organization that complement traditional symptom-based psychiatric assessment.

We emphasize that the present work should be interpreted as a case study rather than a definitive clinical characterization. The cohort size and diagnostic heterogeneity preclude strong claims about generalizability. Instead, our results demonstrate that Lyapunov-based dynamical analysis of discourse embeddings is sensitive to macroscopic organization of speech in psychosis and robust across embedding models. These findings motivate systematic studies on larger datasets, longitudinal recordings, and cross-linguistic corpora, where dynamical markers may be tested for stability, specificity, and clinical relevance.

Several extensions of this framework are currently being investigated. First, supervised and unsupervised classification approaches will be explored using Lyapunov-derived features (e.g., KY dimension, number of positive exponents, maximal exponent) to distinguish healthy controls from psychotic subjects and to assess separability across multiple severity levels. Second, ongoing analyses examine Lyapunov statistics computed from individual answers within subjects, aiming to resolve intra-subject variability and contextual modulation of speech dynamics. Finally, complementary distance-based and recurrence-based measures will be integrated to assess how static semantic separation and dynamical complexity jointly contribute to the characterization of thought disorder.

\section*{Acknowledgements}

 We acknowledge the support from the Ministry of Science, Technological Development and Innovation of the Republic of Serbia, Grant No.~451-03-33/2026-03/200017, ~451-03-34/2026-03/200124, ~451-03-34/2026-03/200156 and Grant No. eu-repo/grantAgreement/MESTD/inst-2020/200163/RS//. Additional support was provided by the Faculty of Technical Sciences, University of Novi Sad through the project “Scientific and Artistic Research Work of Researchers in Teaching and Associate Positions at the Faculty of Technical Sciences, University of Novi Sad 2026” (No. 01-3609/1), as well as by Institutional funding from the Faculty of Medicine, University of Belgrade, 2024-5 (record no. 451-03-65/2024-03/200110). Furthermore, the authors would like to thank Ksenija Mišić, Sanja Andrić Petrović, Vanja Mandić Maravić, Petar Vuković and Vesna Stefanović for their help in the process of data collection, and Mila Jakšić for her help with the transcript revision process.


\section*{Appendix  I}

\subsection{Clinical characteristics of the psychotic cohort}
\label{app:clinical_cohort}

This appendix summarizes the clinical characteristics of the psychotic cohort analyzed in this study. 
Diagnostic categories and symptom severity scores were assigned by clinicians during routine assessment and are reported here for descriptive purposes only. 
Total PANSS score was used for coarse severity stratification in the main analysis.

Severity groups (Mild, Moderate, Severe) were assigned based on total PANSS score using the thresholds described in the main text. 
These categories are intended for coarse stratification and should not be interpreted as homogeneous clinical states, 
as subjects within the same severity group may differ substantially in symptom composition and diagnostic profile.

\begin{table}[hh]
\centering
\scriptsize
\caption{Clinical characteristics of the psychotic cohort. Diagnoses are reported according to ICD-10.
Severity groups are assigned based on PANSS total score and are used for stratification in the dynamical analysis.}
\label{tab:clinical_cohort}
\begin{tabular}{l  l c c c c c}
\hline
\textbf{ID}  & \textbf{ICD-10 diag.} &
\textbf{PANSS P} & \textbf{PANSS N} & \textbf{PANSS G} &
\textbf{PANSS Total} & \textbf{Severity} \\
P1  & F20.0                & 23 & 31 & 45 & 99  & Severe \\
P2    & F23                  & 15 & 24 & 45 & 84  & Moderate \\
P16  & F20.9                & 31 & 26 & 54 & 111 & Severe \\
P20  & F20.0                & 29 & 27 & 54 & 110 & Severe \\
P21  & F20.1                & 18 & 44 & 49 & 111 & Severe \\
P24  & F29                  & 21 & 26 & 51 & 98  & Severe \\
P26  & F20.0                & 24 & 28 & 56 & 108 & Severe \\
P3   & F29             & 16 & 20 & 46 & 82  & Moderate \\
P4    & F29                  & 19 & 26 & 46 & 91  & Moderate \\
P8   & F29                  & 21 & 29 & 40 & 90  & Moderate \\
P12  & F29                  & 20 & 23 & 47 & 90  & Moderate \\
P14  & F23.9    & 23 & 22 & 43 & 88  & Moderate \\
P15  & F23.0                & 14 & 21 & 40 & 75  & Moderate \\
P17  & F20.0                & 19 & 22 & 39 & 80  & Moderate \\
P18  & F20.0                & 20 & 21 & 41 & 82  & Moderate \\
P19  & F20.0                & 16 & 29 & 46 & 91  & Moderate \\
P5    & F29                  & 13 & 19 & 36 & 68  & Mild \\
P6    & F25.1                & 13 & 22 & 36 & 71  & Mild \\
P7    & F23.0                & 13 & 9  & 27 & 49  & Mild \\
P9   & F20.0                & 14 & 19 & 35 & 68  & Mild \\
P10  & F20.0                & 17 & 17 & 27 & 64  & Mild \\
P11 & F29           & 14 & 20 & 35 & 69  & Mild \\
P22  & F20.3                & 12 & 15 & 28 & 55  & Mild \\
P23  & F20.0                & 12 & 19 & 32 & 63  & Mild \\
P25  & F29                  & 14 & 20 & 39 & 73  & Mild \\
\hline
\end{tabular}
\end{table}

\subsection{Robustness with respect to word embedding strategy}\label{app:word_strategy}

In preliminary analyses, we evaluated multiple word embedding strategies, including
token-level embeddings, sentence-contextualized embeddings, and alternative pooling
schemes. Across all strategies, word-level Lyapunov spectra remained uniformly
contracting and did not exhibit qualitative differences in either the sign structure
or ordering of exponents. These findings indicate that the observed word-level
dynamics are robust to the specific embedding construction. For clarity and
consistency, we therefore focus on sentence-contextualized word embeddings throughout
the main text.

\subsection{Sentence-level Lyapunov statistics as a transitional regime}
\label{app:sentence_lyap}

At the sentence level, Lyapunov spectra exhibit pronounced within-group heterogeneity.
For each clinical group, some subjects display fully contracting spectra with all
Lyapunov exponents negative, while others show a small number of positive exponents.
As a consequence, group-averaged statistics are characterized by large standard
deviations and do not reliably represent a single dynamical regime.

This variability reflects the short temporal and semantic span of individual sentences.
At this scale, nonlinear estimates are highly sensitive to local lexical or syntactic
fluctuations, leading to subject-dependent mixtures of weakly expanding and contracting
dynamics. The resulting bimodal behavior renders group-level averages of sentence-level
Lyapunov features statistically unstable and unsuitable for resolving clinical severity.

\subsection{Lyapunov Exponents and Kaplan--Yorke Dimension Using an Alternative Language Model}\label{app:vidi}

In this Appendix, we report dynamical statistics obtained using an alternative transformer-based language model with 1024-dimensional embedding vectors. The analysis parallels that presented in the main text for BERTić embeddings and serves to assess the robustness of Lyapunov-based speech dynamics with respect to the choice of language model.

 All  subjects were analyzed using identical pipelines for trajectory construction, Lyapunov spectrum estimation, and aggregation of dynamical invariants.

The reported measures include the maximal Lyapunov exponent, the sum of Lyapunov exponents, the number of positive exponents, and the Kaplan--Yorke (KY) dimension, which estimates the effective dimensionality of the reconstructed semantic attractor. Because the dimensionality of the embedding space differs from that used in the main text, absolute numerical values of these quantities are not expected to match across models. Accordingly, the results below are intended for qualitative comparison of group-wise trends and relative ordering rather than direct numerical equivalence.

\subsubsection{Word-level embeddings}

Table~\ref{tab:word_lyap_stats} summarizes Lyapunov statistics obtained from word-level embeddings under the four-group clinical classification. Across all groups, the maximal Lyapunov exponent is negative and the sum of exponents is strongly negative, indicating globally contracting and dissipative dynamics. The number of positive exponents is zero in all cases, implying the absence of sustained high-dimensional chaos at the word scale.

Group-wise differences are small and do not lead to qualitative separation between healthy and psychotic speech. This confirms that word-level embeddings capture local semantic coherence but lack sufficient temporal and contextual structure for robust nonlinear dynamical discrimination.

\begin{table}[h]
\centering
\begin{tabular}{|l|c|c|}
\hline
\textbf{Group} & $<LE_{max}>$ & $<\Sigma>$  \\
\hline
Moderate & $-0.4698 \pm 0.0488$ & $-925.41 \pm 39.72$ \\
Mild     & $-0.4826 \pm 0.0529$ & $-935.69 \pm 31.41$ \\
Healthy  & $-0.5154 \pm 0.0577$ & $-952.09 \pm 27.52$ \\
Severe   & $-0.4342 \pm 0.0432 $ & $-929.14 \pm 33.30$ \\
\hline
\end{tabular}
\caption{Lyapunov statistics for word embeddings using four-group classification.}
\label{tab:word_lyap_stats}
\end{table}

\subsubsection{Answer-level embeddings: two-, three-, and four-group classifications}

In contrast to word-level results, answer embeddings yield positive maximal Lyapunov exponents and high-dimensional attractors across all groups. This confirms that discourse-scale embeddings encode nontrivial nonlinear dynamics with expanding directions in semantic phase space.

Table~\ref{tab:llm_answer_2groups} presents the two-group comparison (Healthy vs.\ Psychosis). Healthy subjects exhibit substantially higher KY dimensions and a markedly larger number of positive Lyapunov exponents than the psychosis group, indicating richer and higher-dimensional semantic dynamics. While the maximal Lyapunov exponent is also higher on average in healthy speech, overlap between groups suggests that this single quantity is insufficient for resolving clinical heterogeneity on its own.

\begin{table*}[t]
\centering
\caption{Lyapunov spectrum statistics for alternative LLM answer embeddings under two-group classification (Healthy vs.\ Psychosis). Values are given as mean $\pm$ standard deviation.}
\label{tab:llm_answer_2groups}
\begin{tabular}{|l|c|c|c|c|}
\hline
Group & $<LE_{max}> $ & $<\Sigma>$ &  $<N_{LE}>$ & $<D_{KY}>$ \\
\hline
Psychosis & $0.241 \pm 0.078$ & $-337.827 \pm 30.824$ & $51.38 \pm 35.74$ & $126.90 \pm 85.10$ \\
Healthy   & $0.353 \pm 0.096$ & $-318.811 \pm 39.555$ & $120.78 \pm 58.37$ & $280.28 \pm 125.46$ \\
\hline
\end{tabular}
\end{table*}

Tables~\ref{tab:llm_answer_3groups} and~\ref{tab:llm_answer_4groups} report results under three- and four-group clinical stratifications, respectively. Across both schemes, healthy subjects consistently exhibit the highest KY dimensions and numbers of positive Lyapunov exponents. Psychotic subgroups show reduced dimensionality and substantial overlap, with no clear monotonic dependence on clinical severity.

\begin{table*}[t]
\centering
\caption{Lyapunov spectrum statistics for alternative LLM answer embeddings under three-group classification.}
\label{tab:llm_answer_3groups}
\begin{tabular}{|l|c|c|c|c|}
\hline
Group & $<LE_{max}> $ & $<\Sigma>$ &  $<N_{LE}>$ & $<D_{KY}>$ \\
\hline
Severe   & $0.214 \pm 0.065$ & $-317.023 \pm 22.146$ & $55.17 \pm 37.23$ & $133.70 \pm 86.59$ \\
Mild+Moderate    & $0.250 \pm 0.082$ & $-346.393 \pm 29.796$ & $50.11 \pm 36.25$ & $124.64 \pm 97.64$ \\
Healthy  & $0.353 \pm 0.096$ & $-318.811 \pm 39.555$ & $120.78 \pm 58.37$ & $280.28 \pm 125.46$ \\
\hline
\end{tabular}
\end{table*}

\begin{table*}[t]
\centering
\caption{Lyapunov spectrum statistics for alternative LLM answer embeddings under four-group clinical classification.}
\label{tab:llm_answer_4groups}
\begin{tabular}{|l|c|c|c|c|}
\hline
Group & $<LE_{max}> $ & $<\Sigma>$ &  $<N_{LE}>$ & $<D_{KY}>$ \\
\hline
Severe   & $0.214 \pm 0.065$ & $-317.023 \pm 22.146$ & $55.17 \pm 37.23$ & $133.69 \pm 86.59$ \\
Moderate & $0.250 \pm 0.099$ & $-349.820 \pm 20.819$ & $45.67 \pm 30.31$ & $132.82 \pm 101.23$ \\
Mild     & $0.251 \pm 0.067$ & $-343.346 \pm 35.663$ & $54.55 \pm 42.77$ & $116.45 \pm 75.43$ \\
Healthy  & $0.353 \pm 0.096$ & $-318.811 \pm 39.555$ & $120.78 \pm 58.37$ & $280.28 \pm 125.46$ \\
\hline
\end{tabular}
\end{table*}

\subsubsection{Robustness across language models}

Comparing these results with those obtained using BERTić embeddings in the main text reveals strong qualitative agreement. In both models, answer embeddings provide the clearest separation between healthy and psychotic speech, driven primarily by differences in attractor dimensionality and the multiplicity of unstable directions. Word embeddings remain dynamically contracting and weakly discriminative across models.

Although the alternative model tends to yield slightly higher absolute KY dimensions and numbers of positive exponents—particularly in healthy subjects—the relative ordering of clinical groups and the dominance of discourse-level features are preserved. Because the length of the Lyapunov spectrum depends explicitly on embedding dimensionality, component-wise comparison of individual exponents across models is not meaningful. Instead, the observed consistency of aggregate invariants supports the interpretation that Lyapunov-based measures capture model-robust structural properties of speech dynamics.

Taken together, the Appendix results confirm that nonlinear dynamical markers derived from answer embeddings are robust to moderate variation in the underlying language model. Healthy discourse consistently exhibits higher-dimensional, more flexible semantic dynamics, whereas psychotic speech—although nonlinear—evolves within lower-dimensional and more constrained attractor structures. This robustness strengthens the case for Lyapunov-based invariants as candidate biomarkers of discourse-level disorganization in psychosis.

\section*{Appendix II}

\subsection*{Static semantic distance analysis of question--answer and answer--answer embeddings}

To complement the dynamical analysis based on Lyapunov spectra, we performed a static semantic distance analysis on speech embeddings derived from structured clinical interviews. This appendix reports group-level statistics for (i) paired Question--Answer (QA) distances, probing local semantic coherence between interviewer prompts and patient responses, and (ii) Answer--Answer (AA) distances, characterizing the internal semantic dispersion of patient discourse across successive answers. Distances were computed using six commonly employed measures: cosine distance, Euclidean distance \cite{CoverThomas2006}, Kullback--Leibler (KL) divergence \cite{Kullback1951}, Hellinger distance \cite{Hellinger1909}, Bhattacharyya distance \cite{Bhattacharyya1943}, and Jensen--Shannon (JS) divergence \cite{Lin1991JS}.

\subsection*{A. Question--Answer semantic distances}

Table~\ref{tab:QA_static} summarizes the mean $\pm$ standard deviation of QA distances across diagnostic groups. Distances were first averaged over all question--answer pairs within each subject and subsequently aggregated across subjects within each group.

\begin{table*}[t]
\centering
\caption{Group-level statistics of QA semantic distances (mean $\pm$ SD across subjects).}
\label{tab:QA_static}
\begin{tabular}{|l|c|c|c|c|c|c|}
\hline
Group & Cosine & Euclidean & KL & Hellinger & Bhattacharyya & JS \\
\hline
Severe Psychosis & 0.0557 $\pm$ 0.0083 & 4.482 $\pm$ 0.343 & 0.165 $\pm$ 0.090 & 0.200 $\pm$ 0.057 & 0.0616 $\pm$ 0.0337 & 0.0503 $\pm$ 0.0263 \\
Moderate Psychosis & 0.0508 $\pm$ 0.0054 & 4.276 $\pm$ 0.353 & 0.137 $\pm$ 0.079 & 0.178 $\pm$ 0.057 & 0.0507 $\pm$ 0.0320 & 0.0412 $\pm$ 0.0245 \\
Mild Psychosis  & 0.0537 $\pm$ 0.0067 & 4.432 $\pm$ 0.330 & 0.167 $\pm$ 0.075 & 0.205 $\pm$ 0.056 & 0.0622 $\pm$ 0.0281 & 0.0514 $\pm$ 0.0225 \\
Healthy & 0.0506 $\pm$ 0.0052 & 4.110 $\pm$ 0.228 & 0.133 $\pm$ 0.042 & 0.184 $\pm$ 0.036 & 0.0480 $\pm$ 0.0171 & 0.0407 $\pm$ 0.0135 \\
\hline
\end{tabular}

\centering
\caption{Group-level statistics of AA semantic distances (mean $\pm$ SD across subjects).}
\label{tab:AA_static}
\begin{tabular}{|l|c|c|c|c|c|c|}
\hline
Group & Cosine & Euclidean & KL & Hellinger & Bhattacharyya & JS \\
\hline
Severe Psychosis  & 0.0529 $\pm$ 0.0076 & 3.519 $\pm$ 0.269 & 0.182 $\pm$ 0.109 & 0.148 $\pm$ 0.034 & 0.0367 $\pm$ 0.0185 & 0.0314 $\pm$ 0.0144 \\
Moderate Psychosis  & 0.0460 $\pm$ 0.0099 & 3.393 $\pm$ 0.567 & 0.154 $\pm$ 0.082 & 0.137 $\pm$ 0.031 & 0.0323 $\pm$ 0.0147 & 0.0274 $\pm$ 0.0115 \\
Mild Psychosis  & 0.0491 $\pm$ 0.0123 & 3.524 $\pm$ 0.755 & 0.194 $\pm$ 0.154 & 0.147 $\pm$ 0.039 & 0.0376 $\pm$ 0.0209 & 0.0310 $\pm$ 0.0152 \\
Healthy & 0.0437 $\pm$ 0.0127 & 3.171 $\pm$ 0.604 & 0.138 $\pm$ 0.075 & 0.130 $\pm$ 0.026 & 0.0274 $\pm$ 0.0108 & 0.0239 $\pm$ 0.0085 \\
\hline
\end{tabular}
\end{table*}

Across all metrics, healthy subjects tend to exhibit smaller QA distances, consistent with stronger semantic alignment between questions and answers. Psychotic groups show modestly increased distances, with no clear monotonic dependence on clinical severity. One-way ANOVA and Kruskal--Wallis tests reveal that only the Euclidean distance approaches statistical significance, while all other measures remain non-significant, reflecting substantial overlap between groups.

\subsection*{B. Answer--Answer semantic distances}

Table~\ref{tab:AA_static} reports group-level statistics of AA distances, computed from all pairwise distances between answers within each subject (upper triangle of the distance matrix), followed by aggregation across subjects within each diagnostic group.

Healthy subjects consistently display the smallest AA distances, indicating tighter clustering of answers in embedding space and greater internal semantic consistency across responses. Psychotic groups show increased dispersion, particularly in severe and mild psychosis. Nevertheless, none of the AA distance measures yield statistically significant group separation.

\subsection*{C. Relation to dynamical analysis}

Overall, the static distance analyses reveal only weak group-level trends and fail to robustly distinguish healthy from psychotic speech. The substantial overlap between groups across all static metrics indicates that semantic proximity alone—when temporal ordering and trajectory structure are ignored—is insufficient to characterize clinically relevant alterations in discourse organization.

These findings provide an important baseline for the dynamical framework developed in the main text. In contrast to static measures, Lyapunov spectra and derived invariants capture the temporal organization, instability structure, and effective dimensionality of speech trajectories, yielding a markedly clearer separation between healthy and psychotic discourse. The static results reported here therefore serve not as competing evidence, but as a complementary reference highlighting the added value of a dynamical systems approach to language analysis.

\bibliography{sample}

\end{document}